\documentclass[journal]{IEEEtran}
\usepackage[cmex10]{amsmath}
\usepackage{latexsym,amsmath,enumerate,amssymb,amsbsy}
\usepackage{supertabular}
\usepackage{graphics,color}
\newtheorem{theorem}{Theorem}[section]
\newtheorem{lemma}[theorem]{Lemma}

\newtheorem{remark}[theorem]{Remark}
\newtheorem{proposition}[theorem]{Proposition}
\newtheorem{corollary}[theorem]{Corollary}
\numberwithin{equation}{section}
\newtheorem{example}[theorem]{Example}

\def\0{{\mathbf{0}}}
\def\1{{\mathbf{1}}}

\numberwithin{equation}{section}

\title{Checkable    Codes from  Group Rings}
\author{Somphong~Jitman,~\IEEEmembership{Student Member,~IEEE}, \linebreak[3]%
        San~Ling,~\IEEEmembership{}\linebreak[3]%
        Hongwei~Liu,~\IEEEmembership{}\linebreak[3]%
        and~Xiaoli~Xie~\IEEEmembership{}
\thanks{S.~Jitman, and S.~Ling are with the Division of
  Mathematical Sciences,
  School of Physical and Mathematical Sciences, Nanyang Technological
  University,  21 Nanyang Link, Singapore 637371,
  Republic of Singapore
  (emails: pu738241@e.ntu.edu.sg, lingsan@ntu.edu.sg).}%
\thanks{S.~Jitman is also with the Department of Mathematics,
  Faculty of Science, Chulalongkorn University,
  Bangkok 10330, Thailand.}%
  \thanks{H.~Liu, and X.~Xie are with the Department of Mathematics, Huazhong Normal University,
  Wuhan, Hubei 430079, China
  (emails:\{h\_w\_liu,xiexiaoli\_{1985}\}@yahoo.com.cn).}%
\thanks{The work of S.~Jitman and S.~Ling was  partially supported by the Singapore Ministry of Education under
Research Grant T208B2206. The work of H.~Liu and X.~Xie  was done under the National Natural Science Foundation of China, Grant No. 10871079.}%
}

\begin{document}

\maketitle

\begin{abstract}
We study  codes  with a single check element derived from  group rings, namely, checkable codes. The notion of a
code-checkable group ring is introduced.  Necessary and sufficient conditions for a group ring to be
code-checkable are given in the case where the group is a finite abelian group and the ring is a finite field.
This characterization leads to many good examples, among which two checkable codes and two shortened codes have
minimum distance better than the lower bound given in Grassl's online table. Furthermore, when a group ring is
code-checkable, it is shown that every code in such a group ring admits a generator, and that its dual is also
generated by an element which may be deduced directly from a check element of the original code. These are
analogous to the generator and parity-check polynomials of cyclic codes. In addition, the structures of
reversible and complementary dual checkable codes are established as generalizations of reversible  and
complementary dual cyclic codes.

\end{abstract}

\begin{keywords} checkable code, group ring,   Sylow $p$-subgroup, zero-divisor code, reversible code, complementary dual code.
\end{keywords}

\section{Introduction}\label{sec:intro}
 A group ring code  is originally defined to be an
ideal in the group ring $\mathbf{F}G$, where $\mathbf{F}$ is a finite field and $G$ is a finite group. When $G$
is cyclic, this concept characterizes the classical cyclic codes over $\mathbf{F}$. In general, when $G$ is
abelian, they are called abelian codes and have been studied by many authors (see \cite{Be1967}-\cite{Be1967-2},
\cite{Mac1969}-\cite{Mac1970}, and \cite{Ch1992}).

Recently, new techniques for constructing codes have been established for an arbitrary group ring $RG$ in
\cite{HuHu2007}, where $R$ is an associative ring with identity $1\ne 0$ and $G$ is a finite group. For a
submodule $W$ of the $R$-module $RG$ and a zero-divisor $u$  in $RG$, a {\it zero-divisor code} generated by $u$
relative to $W$ is defined to be $\mathcal {C}:=\{wu\mid w\in W\}=Wu$.  Many existing codes coincide with
special types of zero-divisor codes (cf. \cite{HuHu2007}-\cite{Hurley-H}, and \cite{McHu}).

One of the most  interesting is a zero-divisor code determined by a single check element, i.e., there exists
$v$ in $RG$ such that $\mathcal{C}=Wu=\{y\in RG\mid yv=0\}$. Such a code is  called a {\it checkable code} and
the element $v$ is called a {\it check element}. A group ring  is said to be  {\it code-checkable} if all its
non-trivial ideals are checkable codes. These codes are of interest since they can be viewed as a generalization
of the classical cyclic codes.
 For   a finite field $\mathbf{F}$ and a cyclic group $G$ of order $n$, $\mathbf{F}G\cong \mathbf{F}[X]/\langle X^n-1 \rangle$ is a principal ideal ring, where $\mathbf{F}[X]$ is
the ring of polynomials over $\mathbf{F}$.  All ideals  of $\mathbf{F}G$  are cyclic codes. Every non-trivial
ideal  is checkable, where the ideal is generated by the generator polynomial and the reciprocal polynomial of
the parity-check polynomial acts  as  a check element. Therefore, $\mathbf{F}G$ is code-checkable.

We extend this study to the group ring $\mathbf{F}G$, where $\mathbf{F}$ is a finite field and $G$ is a finite
abelian group. Necessary and sufficient conditions for $\mathbf{F}G$ to be code-checkable are determined. This
characterization allows us to find various examples of  good  codes. Four new codes which have minimum distance
better than the lower bound given   in Grassl's table~\cite{Gr2010} are presented. Many other examples found
also have minimum distance as good as the best known ones in \cite{Gr2010}. Furthermore, it is also shown that,
when $\mathbf{F}G$ is a code-checkable  group ring, every zero-divisor code in $\mathbf{F}G$ is of the form
$\mathbf{F}Gu = \{y\in \mathbf{F}G\mid yv=0\} $ for some $u,v \in \mathbf{F}G$, and that its dual is given by
$\mathbf{F}Gv^{(-1)}$, where $v^{(-1)} $ is defined to be $v^{(-1)} = \sum_{g \in G} v_{g^{-1}} g$ for $v =
\sum_{g \in G} v_g g$. As seen above, when  $G$ is a cyclic group, i.e., in the case of cyclic codes over
$\mathbf{F}$, $u$ and $v^{(-1)}$ may be regarded as the analogs of the generator and parity-check polynomials.
In this sense, the class of codes studied in this paper can be regarded as a generalization of cyclic codes.
Indeed, when $G$ is a finite abelian group, the group ring $\mathbf{F}G$ is isomorphic to  some $\mathbf{F}[X_1,
\ldots , X_t]/\langle X_1^{n_1} -1, \ldots , X_t^{n_t} -1 \rangle$ (cf.~\cite{Di2000}), so the elements $u$ and
$v^{(-1)}$ may be regarded as the multivariate generator and parity-check polynomials of a checkable abelian
code. Moreover, we derive the structures of reversible  and complementary dual checkable codes which may have
application in certain data storage, computing,  and retrieval systems. These codes are  generalizations of
reversible  and complementary dual cyclic codes (cf. \cite{AbGh2006},   \cite{Ma1964}, and \cite{YaMa1994}).

The paper is organized as follows. Some basic concepts and   necessary terminologies are introduced in
Section~\ref{sec:prelim}. In Section~\ref{sec:char},  we present a characterization of code-checkable group
rings together with some related properties.  We provide structural characterizations of reversible  and
complementary dual checkable codes    in Section \ref{sec:spec}.  In Section~\ref{sec:examples},  some examples
from the family of checkable codes and their modifications are discussed, including four  new codes and numerous
good codes. Finally, we conclude with a summary of results in Section~\ref{sec:conclu}.

\section{Preliminaries}\label{sec:prelim}
In order for the exposition in this paper to be self-contained, we introduce some basic concepts and necessary
terminologies used later in this paper. The readers may find further details in \cite{Du2004}-\cite{FiSe1976},
\cite{HuHu2007}-\cite{Hurley-H},     and  \cite {Milies-Sehgal}.

\subsection{Groups and Group Rings}

 Let $G$ be a finite   group   and $p$ a prime number. If $G$ is of order $p^am$, where $a$ is a
non-negative integer and $m$ is a positive integer  such that $p\nmid m$, then a subgroup of order $p^a$ is
called a {\em Sylow $p$-subgroup} of $G$.

Throughout, we assume that $G$ is abelian of order $n$, written multiplicatively  (with identity $1$). Let
$\mathbf{F}$ denote a finite field of characteristic $p$ and denote by  $\mathbf{F}G$ the {\it group ring} of
$G$ over~$\mathbf{F}$. The elements in $ \mathbf{F}G$ will be written as $\sum\limits_{g\in G}\alpha_{{g }}g $,
where $ \alpha_{g }\in \mathbf{F} $, and the addition and the multiplication are given by
$$
\sum\limits_{g\in G}  \alpha_g{g} +\sum\limits_{g\in G}   \beta_g{g} :=\sum\limits_{g\in G} (\alpha_g+\beta_g)g
$$ and $$ \left(\sum\limits_{g\in G}  \alpha_g{g}\right)\hspace{-.25cm}\left( \sum\limits_{h\in G}
\beta_h{h}\right):=\sum\limits_{g,h \in G} (\alpha_g\beta_h)gh.
$$
Obviously, $\mathbf{F}G$ is an $\mathbf{F}$-vector space with a basis $G$, where the scalar multiplication is
defined by $$r\sum\limits_{g\in G} \alpha_g{g}:=\sum\limits_{g\in G} (r\alpha_g){g},$$ for all $r\in \mathbf{F}$
and $ \sum\limits_{g\in G} \alpha_g{g}\in \mathbf{F}G$. As $G$ is abelian, the group ring $\mathbf{F}G$ is
commutative.

 Let  $\{g_1,g_2,\dots,g_n\}$ be a fixed list of the elements in $G$ and $M_{n}(\mathbf{F})$ denote the ring
of~$n\times n$ matrices over $\mathbf{F}$. For $u=\sum\limits_{i=1}^{n}u_{g_{i}} g_{i}\in \mathbf{F}G$, let
 $U\in M_{n}(\mathbf{F})$  be defined by
\begin{equation}\label{eq-U}
U=\left(\begin{array}{cccc}
u_{g^{-1}_1g_{1}}&u_{g_{1}^{-1}g_{2}}&\cdots&u_{g_{1}^{-1}g_{n}}\\
u_{g_{2}^{-1}g_{1}}&u_{g_{2}^{-1}g_{2}}&\cdots&u_{g_{2}^{-1}g_{n}}\\
\vdots&\vdots&\ddots&\vdots\\
u_{g_{n}^{-1}g_{1}}&u_{g_{n}^{-1}g_{2}}&\cdots&u_{g_{n}^{-1}g_{n}}
\end{array}
\right).
\end{equation}
The map
 $\tau : \mathbf{F}G \rightarrow
M_{n}(\mathbf{F})$  given by \[u  \mapsto U^T,\] where $U^T$ is the transpose of $U$, is well-known as a {\it
left regular representation} of $\mathbf{F}G$ (cf. \cite[Chapter 2]{Fa2001}, and \cite[Example
4.1.6]{Milies-Sehgal}). This representation plays a vital role in studying the generator and parity-check
matrices of codes mentioned later.

 An   element $a\in \mathbf{F}G$ is called    a {\it unit} if there exists $  b\in \mathbf{F}G$ such that
$ab=1$. A  non-zero element $u\in \mathbf{F}G$ is called    a {\it zero-divisor} if there exists $0\ne v\in
\mathbf{F}G$ such that $uv=0$. For a non-empty subset $S$ of $\mathbf{F}G$, the {\it annihilator} of $S$ is
defined to be   {\it $Ann(S)=\{x\in \mathbf{F}G \mid xs=0, \textnormal{ for all } s\in S\}$}. Note that $Ann(S)$
is an ideal of $\mathbf{F}G$. When $S=\{s\}$, we simply denote by $Ann(s)$ the annihilator $Ann(S)$. An ideal
$I$ of $\mathbf{F}G$ is said to be {\it non-trivial} if $\{0\}\subsetneq I \subsetneq \mathbf{F}G$ and it  is
said to be {\it principal} if it is generated by a single element. We say that $\mathbf{F}G$ is a {\it principal
ideal ring (PIR)} if every ideal of $\mathbf{F}G$ is principal.

In the light of the main result in \cite{FiSe1976}, a characterization of principal ideal group rings is given
as follows.
\begin{theorem}[\cite{FiSe1976}]\label{principalFG}
 Let $G$ be a finite abelian group and $\mathbf{F}$ a finite field of characteristic $p$. Then $\mathbf{F}G$ is
 a PIR if and only if a Sylow $p$-subgroup  of $G$ is cyclic.
\end{theorem}

\subsection{Codes from Group Rings}

A zero-divisor code  has been  introduced for arbitrary group rings in \cite{Hurley-H}. We recall this concept
for a commutative group ring $\mathbf{F}G$ as follows:

Let  $W$ be a subspace of the $\mathbf{F}$-vector space $\mathbf{F}G$ and let $u$ be a zero-divisor in
$\mathbf{F}G$.  The   {\it zero-divisor code} $\mathcal{C}$   generated by $u$ relative to $W$  is defined to be
$\mathcal {C}:=\{wu\mid w\in W\}=Wu$.  The element $u$ is called a {\it generator element} for $\mathcal{C}$.

Given a zero-divisor code $\mathcal {C}=Wu$, then there exists $ 0\ne v\in \mathbf{F}G$ such that $uv=0$ and
hence $cv=0$ for all $c\in \mathcal {C}$. If there is an element $v \in \mathbf{F}G$ such that $\mathcal
{C}=\{y\in \mathbf{F}G\mid yv=0\}=Ann(v)$, the code $\mathcal{C}$ is said to be {\it checkable} and the element
$v$ is called a {\it check element} of $\mathcal {C}$. We note that a check element for a code does not need to
be unique. The group ring $\mathbf{F}G$ is said to be {\it code-checkable} if every non-trivial ideal of
$\mathbf{F}G$ is a checkable code.

Let $u$ be a zero-divisor in $\mathbf{F}G$ and $U$  its  corresponding matrix  defined in (\ref{eq-U}). Assume
that $W$ is a subspace of $\mathbf{F}G$ with a basis $S\subseteq G$ such that $Su$ is linearly independent. If
$|S|=k$, then ${\rm rank}(U)=k$ if and only if the code
  $\mathcal
{C}=Wu $ is an ideal of $\mathbf{F}G$, equivalently, $\mathcal{C}=\mathbf{F}Gu$ (see
\cite[Theorem~7.2]{Hurley-H}).

To determine whether $\mathbf{F}G$ is code-checkable, it suffices to consider all zero-divisor codes $\mathcal
{C}$ where $\mathcal {C}=\mathbf{F}Gu$. From this characterization, a generator matrix for $\mathcal{C}$ can be
defined to be any $k$ linearly independent rows of $U$.

A zero-divisor $u\in \mathbf{F}G$  is called  {\it principal} if there exists $0\ne v\in \mathbf{F}G$ such that
$uv=0$ and ${\rm rank}(V)=n-{\rm rank}(U)$, where $U$ and $V$ are the corresponding matrices of $u$ and $v$,
respectively. The following characterization is proved in~\cite{Hurley-H}.
\begin{lemma}[{\cite[Corollary 4.1]{Hurley-H}}]
Let $u$ be a zero-divisor in $ \mathbf{F}G$. Then the zero-divisor code $\mathbf{F}Gu$ is checkable if and only
if $u$ is principal.
\end{lemma}
In this case, it is easy to see that the corresponding element $v$ is a check element of $\mathcal{C}$. As
$uv=0$, it follows that $UV=0$. Hence, by the rank condition, a parity-check matrix for $\mathcal{C}$  can be
defined to be any $n-k$ linearly independent columns~of~$V$.

For $a=\sum\limits_{g\in G}a_g g$ and $b=\sum\limits_{g\in G}b_g g$  in $\mathbf{F}G$, let $\langle a,b \rangle$
denote the {\it Euclidean inner product} of the coefficient vectors of $a$ and $b$, i.e., \[ \langle a,b
\rangle= \sum_{g \in G}a_g b_g.\] For a code $\mathcal{C}\subseteq \mathbf{F}G$, the {\it dual code}
$\mathcal{C}^\perp$ of $\mathcal{C}$ is defined by \[\mathcal{C}^\perp=\{a \in \mathbf{F}G \mid \langle a,c
\rangle= 0 \textnormal{ for all } c \in \mathcal{C}\}.\]

\section{Checkable Codes and Code-Checkable Group Rings}\label{sec:char}
In this section, we present  the main results of this paper. A characterization of code-checkable group rings
and some relevant properties are given.

\begin{proposition}\label{lemma-1} Let $\mathbf{F}$ be a finite field and $G$ a finite abelian group. Then $\mathbf{F}G$ is   code-checkable
if and only if it is a PIR.
\end{proposition}
\begin{IEEEproof}
Assume that  $\mathbf{F}G$ is  code-checkable. Let $I$ be an arbitrary  ideal of $\mathbf{F}G$. If $I$ is
$\{0\}$ or $\mathbf{F}G$, it is principal. Assume that $I$ is non-trivial. Then there exists a zero-divisor $
v\in \mathbf{F}G$ such that $I=Ann(v)$. Then
\begin{equation}\label{cong}
\mathbf{F}G/Ann(v)\cong \mathbf{F}Gv.
\end{equation}
Next, we show $I$ is principal. Since $\{0\}\subsetneq \mathbf{F}Gv\subsetneq \mathbf{F}G$,   there exists
$0\neq u\in \mathbf{F}G$  such that $\mathbf{F}Gv=Ann(u)$. We claim that $\mathbf{F}Gu=Ann(v)$. It is clear that
$\mathbf{F}Gu\subseteq Ann(v)$. By (\ref{cong}), we have that
$$
|\mathbf{F}G/Ann(v)|=|\mathbf{F}Gv| \,\, \mbox{and}  \,\,|\mathbf{F}G/Ann(u)|=|\mathbf{F}Gu|.
$$
Since $\mathbf{F}G$ is finite, it follows that
$$
|\mathbf{F}Gu|=|\mathbf{F}G|/|Ann(u)|=|\mathbf{F}G|/|\mathbf{F}Gv|=|Ann(v)|.
$$
Hence, $I=Ann(v)=\mathbf{F}Gu$. Therefore,   $\mathbf{F}G$ is a PIR.

Conversely, assume that $\mathbf{F}G$ is a PIR. Let $\mathfrak{J}$ denote the set of all non-trivial ideals of
$\mathbf{F}G$. From the finiteness of $\mathbf{F}G$, it follows that $|\mathfrak{J}|$ is finite. Let $\sigma:
\mathfrak{J}\to \mathfrak{J}$  be defined by
$$
  \ \mathbf{F}Ga\mapsto Ann(a).
$$
Clearly, for each $c\in \mathbf{F}G$, we have $Ann(\mathbf{F}Gc)=Ann(c)$. Hence, if $\mathbf{F}Ga=\mathbf{F}Gb$,
then
$$
Ann(a)=Ann(\mathbf{F}Ga)=Ann(\mathbf{F}Gb)=Ann(b).
$$
This implies that the mapping $\sigma$   is well-defined.

To show that  $\sigma$ is   injective, assume that $\sigma(\mathbf{F}Ga)=\sigma(\mathbf{F}Gb)$, i.e.,
$Ann(a)=Ann(b)$. Since $\mathbf{F}G$ is a PIR, there exists   $0\ne v\in \mathbf{F}G$  such that
$Ann(a)=Ann(b)=\mathbf{F}Gv$, and hence $\mathbf{F}Ga=Ann(v)=\mathbf{F}Gb$.

Since $|\mathfrak{J}|$ is finite,  $\sigma$ is bijective. This implies that every non-trivial ideal   of
$\mathbf{F}G$ is
  a checkable code.
\end{IEEEproof}

A characterization of code-checkable group rings  follows immediately from Theorem~\ref{principalFG} and
Proposition~\ref{lemma-1}.
\begin{theorem}\label{checkable}
 Let $G$ be a finite abelian group  and   $\mathbf{F}$    a finite field of characteristic $p$,
 where $p$ is a prime number. Then the group ring $\mathbf{F}G$ is   code-checkable if and only
  if a Sylow $p$-subgroup of $G$ is cyclic.
\end{theorem}

When $\mathbf{F}G$ is a code-checkable   group ring, Proposition~\ref{lemma-1} and its proof also provide a link
between a checkable code in $\mathbf{F}G$ and its dual. The following result is found in \cite[Theorem
4.6]{Hurley-H}. Here, we give an alternative proof.

For $v = \sum\limits_{g \in G} v_g g \in \mathbf{F}G$, we define $v^{(-1)} = \sum\limits_{g \in G} v_{g^{-1}}
g$.

\begin{corollary}\label{dualcode}
 Let $\mathbf{F}G$ be a code-checkable  group ring.
 Every non-trivial ideal in $\mathbf{F}G$ is of the form $\mathbf{F}G u = Ann(v)$, for some $u, v \in \mathbf{F}G$. Its dual code
 is given by $\mathbf{F}G v^{(-1)}$.
\end{corollary}

\begin{IEEEproof}
%
The fact that every non-trivial ideal in $\mathbf{F}G$ is of the form $\mathcal{C} = \mathbf{F}G u = Ann(v)$ is
already shown in the proof of Proposition~\ref{lemma-1}. For such a code $\mathcal{C}$, we now show that
$\mathcal{C}^{\perp} = \mathbf{F}G v^{(-1)}$.

Write $u =
\sum\limits_{g \in G} u_g g$ and $v = \sum\limits_{h \in G} v_h h$. Hence
\[ 0 = uv = \sum_{k \in G} \left( \sum_{g \in G} u_g v_{g^{-1}k}    \right) k, \]
which implies that $\sum\limits_{g \in G} u_g v_{g^{-1}k}  =0$ for all $k \in G$.

The typical element in $\mathbf{F}G u$ is of the form
\[ \left( \sum_{h \in G}
x_h h   \right) \left( \sum_{g \in G} u_g g   \right) = \sum_{k \in G}
\left( \sum_{g \in G} u_g x_{g^{-1}k}   \right) k . \]
We have that
\[ \begin{array}{rl}
    \sum\limits_{k \in G}
\left( \sum\limits_{g \in G} u_g x_{g^{-1}k}   \right) v_{k^{-1}}
 &= \sum\limits_{g \in G} \left( \sum\limits_{k \in G} x_{g^{-1}k} v_{k^{-1}} \right) u_g \\
 &= \sum\limits_{k \in G} \left( \sum\limits_{g \in G} u_g v_{g^{-1}k} \right) x_{k^{-1}} \\
 &= 0.
\end{array}
\]
This shows that $\mathbf{F}Gv^{(-1)} \subseteq \mathcal{C}^{\perp}$.

It is easy to observe that $v \mapsto v^{(-1)}$ induces an isomorphism of groups $\mathbf{F}Gv \cong \mathbf{F}G
v^{(-1)}$. From the proof of Proposition~\ref{lemma-1}, we have that $|\mathbf{F}G|/|\mathbf{F}Gu| =
|\mathbf{F}Gv| = |\mathbf{F}G v^{(-1)}|$. It therefore follows that $\mathbf{F}Gv^{(-1)} = \mathcal{C}^{\perp}$.
 \end{IEEEproof}
\begin{corollary}\label{cor:size} If   $\mathbf{F}Gu$ is   checkable   with a check element  $v$,
then $|\mathbf{F}Gu|=|\mathbf{F}Gu^{(-1)}|$, $|\mathbf{F}Gv|=|\mathbf{F}Gv^{(-1)}|$, and
$|\mathbf{F}G|=|\mathbf{F}Gu|\cdot|\mathbf{F}Gv|$.
\end{corollary}
\begin{IEEEproof}
  It follows immediately  from   Corollary~\ref{dualcode} and its proof.
\end{IEEEproof}

\section{Some Special Types of Checkable Codes}\label{sec:spec}
In this section, we assume that a group ring $\mathbf{F}G$ is code-checkable and study the structure of some
special types of checkable codes which may have application in certain data storage, computing,  and retrieval
systems.

\subsection{Reversible Checkable Codes}

 For an abelian group $G$ of order $n$, let $\mathcal{L}=\{g_1,g_2,\dots, g_n\}$ denote a fixed list of the elements in $G$.
 For $w=\sum\limits_{i=1}^nw_ig_i$,
 the {\rm reverse} of $w$ with respect to $\mathcal{L}$, denote by ${\rm r}_{\mathcal{L}}(w)$,
 is defined to be
${\rm r}_{\mathcal{L}}(w):=\sum\limits_{i=1}^nw_{n+1-i}g_i$.
  A code $\mathcal{C} \subseteq \mathbf{F}G$ is said to be {\it reversible } with respect to $\mathcal{L}$ if ${\rm r}_{\mathcal{L}}(w)\in
\mathcal{C}$ whenever $w \in \mathcal{C}$. If  the list $\mathcal{L}$ satisfies
\begin{align}
k=g_{n-(i-1)}g_i,\label{eq:gngi}
\end{align}  for  some fixed $k\in G$, and for every $i=1,2,\dots,n$,  then $ {\rm r}_{\mathcal{L}}(w)$ is of the form
\begin{align}
 {\rm r}_{\mathcal{L}}(w)&=\sum\limits_{i=1}^nw_{n+1-i}g_i  =\sum\limits_{i=1}^nw_ig_{n+1-i}\notag\\
                        &=\sum\limits_{i=1}^nw_i k g_{i}^{-1}=k\sum\limits_{i=1}^nw_ig_{i}^{-1}=kw^{(-1)},
\end{align}
for all $w\in \mathbf{F}G$.
\begin{example} Let $G=C_{n_1}\times C_{n_2}\times \dots\times  C_{n_r}$
denote a finite abelian group of order $n=n_1n_2\dots n_r$ written as the product of cyclic groups
$C_{n_j}=\langle x_j\rangle$. Define the list  $\{g_1,g_2,\dots, g_n\}$  of $G$   by
\begin{align}g_{1+j_1+n_1j_2+n_1n_2j_3+\dots+n_1n_2\dots n_{r-1}j_r}=x_1^{j_1}x_2^{j_2}\dots x_r^{j_r},\label{eq:gn}\end{align}
where $0\leq j_i <n_i$ for all $1\leq i\leq r$. Then $g_1=1$, the identity of $G$, and $ g_n=g_{n-(i-1)}g_i$ for
all $1\leq i\leq n$. Hence, this list satisfies  (\ref{eq:gngi}), where $k=g_n$. Note that if $G=\langle
x\rangle$ is cyclic of order $n$, the list represents $ \{1,x,x^2,\dots,x^{n-1}\}$ which corresponds to the set
of monomials $\{1,X,X^2 ,\dots, X^{n-1}\}$ in $\mathbf{F}[X]/\langle X^n-1\rangle$.
\end{example}

Throughout this section, we study reversible checkable codes with respect to a list $\mathcal{L}$ satisfying
(\ref{eq:gngi}).

To complete a characterization of reversible checkable codes, we need the following lemmas.
\begin{lemma}[{\cite[Lemma 1.1]{GrSc2000}}]\label{lem:gen-set}
Given $a\in \mathbf{F}G$, then the set of   generators of $\mathbf{F}Ga$ is $\mathcal{U}(\mathbf{F}G)a$, where
$\mathcal{U}(\mathbf{F}G)$ is the set of   units in $\mathbf{F}G$.
\end{lemma}

\begin{lemma}\label{lem:a=fb}
Let $a$ and $b$ be elements in $\mathbf{F}G$. Then
  $\mathbf{F}Ga=\mathbf{F}Gb$ if and only if $a=fb$ for some unit $f$ in $\mathbf{F}G$.
\end{lemma}
\begin{IEEEproof}
  Assume that $\mathbf{F}Ga=\mathbf{F}Gb$. Note that $a$ is a generator of $ \mathbf{F}Gb$. Then, by Lemma \ref{lem:gen-set}, $a\in \mathcal{U}(\mathbf{F}G)b$ which implies that $a=fb$ for some unit $f\in \mathbf{F}G$.

Conversely, assume that $a=fb$ for some unit $f$ in $\mathbf{F}G$. Then $\mathbf{F}Ga=\mathbf{F}Gfb\subseteq
\mathbf{F}Gb= \mathbf{F}Gf^{-1}a \subseteq \mathbf{F}Ga$.
 Therefore, $\mathbf{F}Ga= \mathbf{F}Gb$ as desired.
\end{IEEEproof}

\begin{theorem}\label{thm:char-rev}
Let $\mathcal{L}$ be a fixed list of $G$ satisfying (\ref{eq:gngi}). Let  $\mathbf{F}Gu$    be a  checkable code
with a check element $v$. Then the following statements are equivalent:
\begin{enumerate}[$i)$]
  \item $\mathbf{F}Gu$ is reversible with respect to $\mathcal{L}$.%
  \item $\mathbf{F}Gu=\mathbf{F}Gu^{(-1)}$.%
  \item $u=au^{(-1)}$   for some unit  $a$ in $\mathbf{F}G$.%
  \item $v=b v^{(-1)}$   for some unit  $ b$ in $\mathbf{F}G$.%
  \item $\mathbf{F}Gv=\mathbf{F}Gv^{(-1)}$.%
  \item $\mathbf{F}Gv$ is reversible with respect to $\mathcal{L}$.
\end{enumerate}
\end{theorem}
\begin{IEEEproof} We prove $i)\Rightarrow ii) \Rightarrow iii)\Rightarrow i)$,  $iii) \Leftrightarrow iv)$, and
 $iv)\Rightarrow v)\Rightarrow vi)\Rightarrow iv)$.
To prove $i)\Rightarrow ii)$, assume that $\mathbf{F}Gu$ is reversible with respect to $\mathcal{L}$. Since
$\mathbf{F}G$ contains $1$, $ku^{(-1)}={\rm r}_{\mathcal{L}}(u)\in \mathbf{F}Gu$.  Then $u^{(-1)}=k^{-1}{\rm
r}_{\mathcal{L}}(u)\in \mathbf{F}Gu$, i.e. $\mathbf{F}G u^{(-1)}\subseteq \mathbf{F}Gu$. Since, by Corollary
\ref{cor:size}, they have the same cardinality, we conclude that $\mathbf{F}G u^{(-1)}= \mathbf{F}Gu$.

The proof of  $ii)\Rightarrow iii)$ is  immediate from Lemma \ref{lem:a=fb}.

To prove $iii)\Rightarrow i)$, assume that there exists a unit  $a \in\mathbf{F}G$ such that  $u=au^{(-1)}$. Let
$wu\in \mathbf{F}Gu$. Then
\begin{align*}
{\rm r}_{\mathcal{L}}(wu)   &=k(wu)^{(-1)}=kw^{(-1)}u^{(-1)}\\
                            &= (kw^{(-1)} a^{-1})u\in \mathbf{F}Gu.
\end{align*} This shows that  $\mathbf{F}Gu$ is reversible.

Next, we prove $iii) \Leftrightarrow iv)$.  Assume that  $u=au^{(-1)}$   for some unit  $a$ in $\mathbf{F}G$.
Since $0=uv=au^{(-1)}v=u^{(-1)}(av) $ and $v^{(-1)}$ is a check element of $\mathbf{F}Gu^{(-1)}$, we have $av\in
\mathbf{F}Gv^{(-1)}$.
 As $a$ is a unit, $v\in \mathbf{F}Gv^{(-1)}$.
 Then, by Corollary \ref{cor:size}, $\mathbf{F}Gv= \mathbf{F}Gv^{(-1)}$.
 Therefore, by Lemma \ref{lem:a=fb}, there exists  a unit  $ b$ in $\mathbf{F}G$ such that  $v=b v^{(-1)}$. The converse is proved using similar arguments.

The equivalence $iv)\Rightarrow v)\Rightarrow vi)\Rightarrow iv)$ is proved similar to
   $iii)\Rightarrow ii) \Rightarrow i)\Rightarrow iii)$.
\end{IEEEproof}

\begin{remark}
To verify whether $\mathbf{F}Gu$ is reversible, by the condition $ii)$, it is equivalent to  checking if
$u^{(-1)}\in \mathbf{F}Gu$.
\end{remark}

When $G=\langle x \rangle $, we know that any non-trivial cyclic code corresponds to some checkable code in
 $\mathbf{F}G$.  According to \cite{Ma1964}, a cyclic code is said to be {\em reversible} if
 its corresponding checkable code
is reversible with respect to the list $ \{1,x,x^2,\dots,x^{n-1}\}$.

For a polynomial $f(X)=f_0+f_1X+\dots+X^t\in \mathbf{F}[X]$   with $f_0\ne 0$, the {\em reciprocal polynomial}
of $f(X)$ is defined to be $f^*(X):=f_0^{-1}X^tf(\displaystyle\frac{1}{X})$. The polynomial $f(X)$ is said to be
self-reciprocal if $f(X)=f^*(X)$. Then the following corollary is immediate from Theorem~\ref{thm:char-rev}.

\begin{corollary}[{\cite[Theorem 1]{Ma1964}}]
The cyclic code generated by a monic polynomial $g(X)$ is reversible if and only if $g(X)$ is self-reciprocal.
\end{corollary}

\subsection{Complementary Dual Checkable Codes}

In this subsection, we study the structure of a checkable code   $\mathbf{F}Gu$ with
 $\mathbf{F}Gu\cap(\mathbf{F}Gu)^\perp=\{0\}$, namely, a {\em complementary dual code} (cf. \cite{YaMa1994}).
We focus on the case where the characteristic $p$ of $\mathbf{F}$ does not divide the order $n$ of $G$ which is
a common restriction as in the study of simple root cyclic codes.

 Under this restriction, the group ring $\mathbf{F}G$ is always code-checkable since the Sylow $p$-subgroup of $G$ is trivial.
 Moreover,   $p\nmid n$  if and only if $\mathbf{F}G$ is semi-simple (cf. {\cite[Chapter 2: Theorem 4.2]{Pa1977}}).
See  \cite{Milies-Sehgal} and  \cite{Pa1977} for further details.

We recall a special ideal of $\mathbf{F}G$ which is key to characterizing the structure of complementary dual
checkable codes. An ideal $A$ of $\mathbf{F}G$ is called a {\em nil ideal} if, for each $a\in A$, there exists a
positive integer $r$ such that $a^r=0$. By the finiteness of $\mathbf{F}G$ and \cite[Theorem 2.7.14 and Theorem
2.7.16]{Milies-Sehgal}, the nil ideal characterizes semi-simplicity of $\mathbf{F}G$ as follows.
\begin{lemma}\label{lem:nil}
A finite group ring  $\mathbf{F}G$ is semi-simple if and only if it  has no non-zero nil ideals.
\end{lemma}

\begin{corollary}\label{CucapCv}
If $\mathbf{F}Gu$ is checkable with a check element $v$, then $\mathbf{F}Gu\cap \mathbf{F}Gv=\{0\}$.
\end{corollary}
\begin{IEEEproof}
Let $w\in   \mathbf{F}Gu\cap \mathbf{F}Gv$. Then    $w=au=bv$ for some $a,b\in \mathbf{F}G$. Hence,
$w^2=aubv=(ab)(uv)=0$ which implies that $\mathbf{F}Gu \cap \mathbf{F}Gv$ is a nil ideal. As $\mathbf{F}G$ is
semi-simple, $\mathbf{F}Gu\cap \mathbf{F}Gv =\{0\}$ by Lemma \ref{lem:nil}.
\end{IEEEproof}

\begin{theorem}\label{thm:char-com-dual}
 Let $\mathbf{F}Gu$ be checkable with a check element~$v$ and   $\mathcal{L}$ a list of  $G$ satisfying (\ref{eq:gngi}). Then the following statements are equivalent.
 \begin{enumerate}[$i)$]
 \item $\mathbf{F}Gu$ is a complementary dual code.%
 \item $\mathbf{F}Gu$ is a reversible code with respect to $\mathcal{L}$.  %
 \item $\mathbf{F}Gv$ is a complementary dual code.%
  \end{enumerate}
\end{theorem}
\begin{IEEEproof}
To prove $i) \Rightarrow ii)$, assume that $\mathbf{F}Gu$ is a complementary dual code. Applying Corollary
\ref{dualcode}, we obtain  $\{0\}= \mathbf{F}Gu\cap(\mathbf{F}Gu)^\perp=\mathbf{F}Gu\cap \mathbf{F}Gv^{(-1)}$
which implies $\mathbf{F}G=\mathbf{F}Gu \oplus\mathbf{F}Gv^{(-1)}$. Since, by Corollary \ref{CucapCv},
$\mathbf{F}Gu\cap \mathbf{F}Gv=\{0\}$, we have
\begin{align*}\mathbf{F}Gv&=\mathbf{F}G\cap
\mathbf{F}Gv\\
&=(\mathbf{F}Gu\cap\mathbf{F}Gv)
\oplus(\mathbf{F}Gv^{(-1)}\cap\mathbf{F}Gv)\\
&=\mathbf{F}Gv^{(-1)}\cap\mathbf{F}Gv.
\end{align*}
Thus, $\mathbf{F}Gv\subseteq \mathbf{F}Gv^{(-1)} $. Since, by Corollary \ref{cor:size}, they have the same
cardinality, it follows that $\mathbf{F}Gv= \mathbf{F}Gv^{(-1)}$. Therefore, $\mathbf{F}Gu$ is reversible by
Theorem \ref{thm:char-rev}.

To prove $ii) \Rightarrow i)$, assume that $\mathbf{F}Gu$ is   reversible with respect to $\mathcal{L}$. Let
$w\in \mathbf{F}Gu\cap(\mathbf{F}Gu)^\perp$. Then, by Corollary \ref{dualcode} and Theorem \ref{thm:char-rev},
$w\in \mathbf{F}Gu\cap \mathbf{F}Gv^{(-1)}=\mathbf{F}Gu\cap \mathbf{F}Gv$.  We have $w=0$ by Corollary
\ref{CucapCv}. Therefore, $\mathbf{F}Gu$ is a complementary dual code.

By Theorem \ref{thm:char-rev}, $ii)$ holds if and only if $\mathbf{F}Gv$ is  reversible,  which is  equivalent
to that $\mathbf{F}Gv$ is a complementary dual code.  This proves $ii) \Leftrightarrow iii)$.
\end{IEEEproof}

\begin{corollary}[{\cite[Corollary]{YaMa1994}}]
Let $\mathbf{F} $ be a finite field of characteristic $p$,  and     $n$    a positive integer  such that $p\nmid
n$. Then a cyclic code of length  $n$ over $\mathbf{F}$ is  a complementary dual code if and only if it is
reversible.
\end{corollary}

\section{Examples}\label{sec:examples}
Many different interesting examples arise from the family of checkable  codes from group rings. In this section,
we discuss some of these examples based on Theorem~\ref{checkable}. We show that various {\it Maximum Distance
Separable (MDS) codes},   $[n,k,d]$ linear codes attaining the Singleton bound $d\leq n-k+1$, are checkable.
Moreover, numerous good checkable codes and new codes are illustrated as well.

\subsection{Some MDS Checkable Codes}
Given a  positive integer  $n$,  we show that $[n,1,n]$ and $[n,n-1,2]$ MDS codes can be constructed as
zero-divisor codes. In many cases, they are checkable.
\begin{lemma}\label{lem1}
 Given a finite field $\mathbf{F}$ and a finite abelian group~$G$,
 then the element $\sum\limits_{g \in G}g$ is always a zero-divisor in the group ring $\mathbf{F}G$.
\end{lemma}
\begin{IEEEproof}
This follows since $(1-g')\sum\limits_{g \in G}g =0$, for all $g'\in G\setminus\{1\}$, where $1$ is
the group identity in $G$.
 \end{IEEEproof}
\begin{corollary}\label{cor2}
Given a finite field $\mathbf{F}$ and a finite abelian group $G$ of order $n$, then there exists an $[n,1,n]$
zero-divisor MDS code constructed from the group ring $\mathbf{F}G$.
\end{corollary}
\begin{IEEEproof}
  From Lemma~\ref{lem1}, $\sum\limits_{g \in G}g$ is a zero divisor in $\mathbf{F}G$. It is easy to see that
  the associated  $U$ of $u=\sum\limits_{g \in G}g$ is the all $1$'s $n\times n$-matrix. Therefore, the code generated by
  $u$ is obviously
  $\{\lambda(1 1 \dots 1)\mid \lambda\in \mathbf{F}\}$, an $[n,1,n]$ MDS code over~$\mathbf{F}$.
  \end{IEEEproof}
\begin{corollary}\label{cor3}
Let $\mathbf{F}$ be a finite field of characteristic $p$ and let $G$  be a finite abelian group of order $n$. If
a Sylow $p$-subgroup of $G$ is cyclic, then there exist checkable $[n,1,n]$ and $[n,n-1,2]$ MDS codes from the
group ring $\mathbf{F}G$.
\end{corollary}
\begin{IEEEproof}
  By Corollary~\ref{cor2}, the code $\mathcal{C}$ generated by $\sum\limits_{g \in G}g$ is an $[n,1,n]$ MDS code.
  Assume that a Sylow $p$-subgroup of $G$ is cyclic.
  From   Theorem~\ref{checkable}, it follows that  $\mathcal{C}$ and its dual $\mathcal{C}^\perp$ are checkable.
 Since $\mathcal{C}$ is
MDS, $\mathcal{C}^\perp$ is again MDS with parameters $[n,n-1,2]$.
\end{IEEEproof}

\begin{remark} Since $(\sum\limits_{g \in G}g)^{(-1)}=(\sum\limits_{g \in G}g)$, the  $[n,1,n]$ MDS
code generated by $\sum\limits_{g \in G}g$ and its dual are  reversible by Theorem \ref{thm:char-rev}. Moreover,
if the characteristic of $\mathbf{F}$ does not divide $n$, then, by Theorem \ref{thm:char-com-dual}, they are
complementary dual.
\end{remark}

\subsection{Good  Codes from Code-Checkable Group Rings}

 We illustrate  some good examples of   checkable codes. Let $\mathbf{F}_q$ denote the finite field
 of order $q$ with characteristic $p$ and let $G$ be an abelian group of order $n$.   When  $G$ is a cyclic group, we know that checkable
 codes from the group ring $\mathbf{F}_qG$ are the classical
  cyclic codes. Hence, we consider examples only in the case
  where $G$ is a non-cyclic abelian group such that a Sylow $p$-subgroup of $G$ is cyclic, i.e., $\mathbf{F}G$ is code-checkable.

With the help of the computer algebra system MAGMA \cite{Bo1997}, generator elements, check elements, and the
actual minimum distances of  checkable codes from   $\mathbf{F}_qG$ are computed in many cases for $q\in
\{2,3,4,5\}$ and $G$ is a non-cyclic abelian group decomposed as a product of two cyclic groups. In numerous
cases, the parameters   of these codes are as good as the best known ones in \cite{Gr2010}. We call such codes
{\em good codes}.  In particular, an optimal $[36,28,6]$ code  and a $[72,62,6]$ code over $\mathbf{F}_5$ with
minimum distances improving by $1$ upon \cite{Gr2010} are found. These are called {\em new codes} presented in
the next subsection.

In Tables~\ref{TableF2}-\ref{TableF5}, a group $G=C_r\times C_s$ of order $n=rs$ denotes the product of cyclic
groups $C_r=\langle x\rangle $ and $C_s =\langle y\rangle$. A vector $u=(u_0u_1u_2\dots u_{n-1})\in
\mathbf{F}_q^n$ represents the element $u(x,y)\in \mathbf{F}_qG$ with respect to the list $\mathcal{L}$ defined
in (\ref{eq:gn}), i.e., $u$ is the coefficients of
\[ u(x,y)=\sum\limits_{j=0}^{s-1}\sum\limits_{i=0}^{r-1}u_{jr+i}x^iy^j \text{ in } \mathbf{F}_q{G}.\]

Given positive integers $n$ and   $k$, the minimum distance of the $[n,k,d]$ codes displayed in the tables
achieve the best known distances \cite{Gr2010}, except for the two codes with asterisk in Table~\ref{TableF5},
where
 the distance improves upon that of the best known ones by 1. Based on the characterizations in Section \ref{sec:spec}, the subscripts $_{\rm R}$ and $_{\rm C}$ indicate
 the reversibility and complementary duality of the codes, respectively. To save space, codes with
 small length, $[n,1,n]$ and $[n,n-1,2]$ MDS codes guaranteed by
Corollary~\ref{cor3}, and  codes with minimum distance $2$ will be omitted.

\subsection{New Codes from Code-Checkable Group Rings}\label{sec:newcodes}

A   checkable code is determined by a check element. We give the check elements of the two new checkable codes
in Table~\ref{TableF5}. In addition, generator elements and the standard generator matrices of these codes are
also provided.  Moreover, other two  optimal  codes   with minimum distances improving by $1$ upon \cite{Gr2010}
are found by shortening a new checkable code.

The $[36,28,6]$ code $\mathcal{C}_{36}$ over $\mathbf{F}_5$  in Table~\ref{TableF5} improves the lower bound on
the minimum distance given in \cite{Gr2010} by $1$ and it is optimal. The code $\mathcal{C}_{36}$ derived from
$\mathbf{F}_5(C_6\times C_6)$ is generated by
$$u_{36}=(021242402043131423014123232100132334)$$
with check element
$$v_{36}=(100004000410431304002224330013242110).$$

The standard generator matrix of $\mathcal{C}_{36}$  is given by
\[\mathcal{G}_{36}=\left(%
\begin{array}{cc }
 &3\,2\,3\,0\,4\,4\,0\,4\,0 \\
 &2\,2\,1\,0\,3\,0\,0\,3\,4 \\
 &4\,0\,0\,0\,2\,1\,3\,2\,4 \\
 &0\,4\,0\,0\,1\,3\,2\,4\,3 \\
 &0\,0\,4\,0\,2\,3\,0\,4\,1 \\
 &1\,2\,2\,0\,3\,4\,0\,3\,3 \\
 &1\,2\,0\,0\,2\,4\,1\,1\,2 \\
 &0\,1\,2\,0\,3\,0\,2\,4\,4 \\
 &3\,1\,2\,0\,3\,3\,0\,0\,0 \\
 &3\,4\,2\,0\,2\,2\,2\,2\,0 \\
 &3\,4\,0\,0\,2\,1\,1\,4\,2 \\
 &0\,3\,4\,0\,3\,0\,4\,4\,2 \\
 &3\,3\,0\,0\,2\,2\,2\,4\,1 \\
I_{27} &0\,3\,3\,0\,4\,1\,1\,1\,3 \\
 &2\,4\,2\,0\,0\,2\,4\,1\,3 \\
 &3\,3\,0\,0\,4\,1\,3\,3\,3 \\
 &0\,3\,3\,0\,2\,1\,3\,0\,0 \\
 &2\,4\,2\,0\,3\,3\,2\,1\,0 \\
 &2\,1\,0\,0\,2\,1\,3\,1\,1 \\
 &0\,2\,1\,0\,4\,1\,0\,2\,0 \\
 &4\,3\,0\,0\,1\,1\,3\,1\,2 \\
 &0\,4\,3\,0\,3\,4\,4\,1\,4 \\
 &2\,4\,3\,0\,4\,0\,1\,3\,2 \\
 &2\,1\,3\,0\,1\,3\,4\,2\,1 \\
 &1\,2\,2\,0\,4\,3\,3\,4\,0 \\
 &3\,2\,3\,0\,2\,3\,2\,0\,4 \\
 &2\,2\,1\,0\,4\,4\,0\,1\,1 \\
\boldsymbol{0}_{1\times 27}&0\,0\,0\,1\,1\,1\,1\,1\,1 \\
 \end{array}%
\right).\]

By shortening $\mathcal{C}_{36}$ at the $1st$ position, we obtain a optimal $[35,27,6]$ code over
$\mathbf{F}_5$. Similarly, a optimal $[34,26,6]$ code over $\mathbf{F}_5$ can be obtained by shortening
$\mathcal{C}_{36}$ at the $1st$ and $2nd$ positions. The  minimum distances of these codes are improved by $1$
from the lower bound given in \cite{Gr2010}.


The $[72,62,6]$ code $\mathcal{C}_{72}$ over $\mathbf{F}_5$ in Table~\ref{TableF5} improves the lower bound on
the minimum distance given in \cite{Gr2010} by $1$. The code $\mathcal{C}_{72}$ derived from
$\mathbf{F}_5(C_6\times C_{12})$ is generated by $u_{72}$ with check element $v_{72}$.

\begin{align*}
\hline\\
    ~~~~~~~~~~~~~~~~u_{72}&=(312411232330313143111221222301122414030013401133430420133323011301020100),   ~~~~~~~~~~~~~~~~\\
               v_{72}&=(100000000441004102234010043124424101300211324012401114201004023203011413).
\end{align*}


The standard generator matrix of $\mathcal{C}_{72}$ is given by {\small
\[\mathcal{G}_{72}=\left(%
\begin{array}{cc }
&3\,0\,0\,0\,3\,3\,2\,3\,3\,3\,0\,1\,2\\
&3\,0\,0\,0\,1\,2\,0\,3\,1\,2\,2\,0\,4\\
&3\,0\,0\,0\,4\,4\,0\,3\,3\,2\,0\,3\,1\\
&3\,0\,0\,0\,4\,2\,2\,0\,3\,4\,0\,1\,4\\
&3\,0\,0\,0\,1\,3\,4\,0\,3\,2\,3\,0\,4\\
&3\,0\,0\,0\,3\,1\,4\,2\,1\,0\,2\,2\,0\\
&4\,0\,0\,0\,4\,3\,0\,4\,4\,2\,0\,2\,2\\
&4\,0\,0\,0\,2\,3\,2\,4\,4\,2\,1\,3\,0\\
&4\,0\,0\,0\,4\,2\,1\,4\,2\,0\,2\,3\,3\\
&4\,0\,0\,0\,3\,1\,3\,1\,3\,4\,2\,2\,2\\
&4\,0\,0\,0\,0\,1\,1\,2\,3\,3\,2\,1\,3\\
&4\,0\,0\,0\,3\,2\,2\,1\,1\,0\,0\,2\,0\\
&3\,0\,0\,0\,3\,4\,0\,1\,0\,2\,1\,4\,1\\
&3\,0\,0\,0\,4\,1\,2\,0\,1\,1\,0\,2\,0\\
&3\,0\,0\,0\,1\,3\,3\,1\,3\,0\,0\,0\,0\\
&3\,0\,0\,0\,2\,3\,2\,2\,3\,1\,2\,2\,4\\
&3\,0\,0\,0\,1\,1\,0\,0\,0\,2\,0\,2\,0\\
&3\,0\,0\,0\,4\,4\,4\,4\,0\,1\,0\,1\,3\\
&3\,0\,0\,0\,3\,4\,4\,4\,0\,1\,4\,4\,1\\
&3\,0\,0\,0\,3\,3\,0\,4\,0\,2\,1\,3\,4\\
&3\,0\,0\,0\,4\,1\,1\,3\,4\,1\,0\,2\,4\\
&3\,0\,0\,0\,0\,0\,1\,4\,2\,4\,2\,3\,4\\
&3\,0\,0\,0\,0\,1\,0\,4\,3\,2\,0\,0\,0\\
&3\,0\,0\,0\,4\,3\,4\,4\,4\,4\,0\,1\,1\\
&3\,0\,0\,0\,1\,3\,2\,3\,0\,2\,4\,1\,3\\
&3\,0\,0\,0\,1\,0\,0\,4\,1\,4\,1\,4\,4\\
&3\,0\,0\,0\,4\,4\,3\,3\,4\,2\,2\,2\,0\\
&3\,0\,0\,0\,2\,1\,3\,2\,0\,2\,4\,4\,1\\
I_{59}&3\,0\,0\,0\,2\,4\,0\,3\,4\,3\,4\,1\,3\\
&3\,0\,0\,0\,4\,0\,2\,2\,3\,0\,1\,0\,2\\
&1\,0\,0\,0\,2\,4\,3\,4\,2\,4\,2\,4\,4\\
&1\,0\,0\,0\,1\,2\,1\,0\,1\,1\,4\,0\,4\\
&1\,0\,0\,0\,4\,0\,0\,3\,4\,2\,0\,3\,3\\
&1\,0\,0\,0\,3\,0\,1\,1\,3\,1\,3\,2\,0\\
&1\,0\,0\,0\,4\,2\,3\,4\,0\,4\,0\,2\,0\\
&1\,0\,0\,0\,1\,4\,4\,1\,1\,4\,4\,3\,2\\
&0\,0\,0\,0\,1\,3\,2\,2\,2\,0\,3\,3\,4\\
&0\,0\,0\,0\,2\,2\,2\,1\,0\,0\,1\,2\,0\\
&0\,0\,0\,0\,2\,3\,1\,2\,4\,3\,1\,0\,4\\
&0\,0\,0\,0\,1\,0\,0\,0\,1\,3\,1\,3\,1\\
&0\,0\,0\,0\,0\,1\,0\,1\,0\,1\,3\,1\,3\\
&0\,0\,0\,0\,0\,0\,1\,3\,1\,0\,1\,3\,1\\
&3\,0\,0\,0\,0\,1\,2\,0\,0\,2\,0\,0\,1\\
&3\,0\,0\,0\,1\,4\,3\,2\,3\,4\,1\,0\,3\\
&3\,0\,0\,0\,2\,3\,3\,0\,4\,1\,1\,3\,4\\
&3\,0\,0\,0\,2\,4\,2\,1\,2\,2\,3\,3\,2\\
&3\,0\,0\,0\,1\,1\,1\,3\,4\,1\,1\,3\,1\\
&3\,0\,0\,0\,0\,2\,1\,1\,2\,4\,2\,4\,0\\
&1\,0\,0\,0\,4\,1\,4\,2\,3\,1\,3\,3\,0\\
&1\,0\,0\,0\,2\,2\,0\,2\,3\,1\,4\,3\,4\\
&1\,0\,0\,0\,3\,3\,3\,2\,2\,0\,2\,1\,0\\
&1\,0\,0\,0\,1\,3\,0\,1\,4\,1\,0\,0\,1\\
&1\,0\,0\,0\,3\,2\,4\,4\,1\,1\,2\,2\,2\\
&1\,0\,0\,0\,2\,1\,1\,4\,0\,4\,4\,2\,3\\
&4\,0\,0\,0\,2\,4\,4\,2\,0\,3\,4\,3\,3\\
&4\,0\,0\,0\,1\,3\,1\,4\,3\,4\,4\,0\,0\\
&4\,0\,0\,0\,3\,3\,4\,3\,3\,0\,1\,4\,4\\
&4\,0\,0\,0\,1\,4\,0\,0\,4\,2\,1\,2\,1\\
&4\,0\,0\,0\,2\,0\,3\,3\,0\,2\,1\,4\,0\\
 &0\,1\,0\,0\,4\,2\,3\,4\,1\,1\,2\,3\,4\\
\boldsymbol{0}_{3\times 59}&0\,0\,1\,0\,3\,3\,3\,2\,1\,3\,0\,3\,1\\
 &0\,0\,0\,1\,3\,2\,4\,4\,4\,3\,2\,1\,1\\
\end{array}%
\right).\]}

\begin{table*}\centering
\caption{Good Checkable Codes from $\mathbf{F}_2G$ } \label{TableF2}
\begin{tabular}[!hbt]{| l|l|l|l|}
  \hline
   $n$& Code $\mathcal{C}$      & Group $G$     &~~~~~~~~~~~~~~~~~Generator Element $u$  and Check Element $v$\\\hline\hline



 $25$&$[25,16,4]_{\rm R,C}$     &$C_{5}\times C_{5}$    &$u=(0111000101010111000101010)$,\\&&& $v=(1000000001101010111010101)$\\
 $ $&$[25,17,4]_{\rm R,C}$     &$C_{5}\times C_{5}$    &$u=(1110001101100000111000100)$,\\&&& $v=(1000000001000100010010111)$\\
\hline

 $27$&$[27,18,4]_{\rm R,C}$     &$C_{3}\times C_{9}$    &$u=(011100010001110110000111011)$,\\&&& $v=(010001001001111111111010010)$\\
\hline

 $45 $&$[45,28,8]$     &$C_{3}\times C_{15}$   &$u=(011101000001100011101100111101011000010000001)$,\\&&&
 $v=(110000000000000011001010000001010111000010111)$\\
 $ $&$[45,29,7]$     &$C_{3}\times C_{15}$   &$u=(001010010100011101100110011101000100110011111)$,\\&&&
 $v=(100000000000000010001100101111101111011100100)$\\
 $ $&$[45,31,6]$     &$C_{3}\times C_{15}$   &$u=(010101110110100011110110001010110010100110001)$,\\&&&
 $v=(110000000000010011001010000110111100110111000)$\\
 $ $&$[45,32,6]$     &$C_{3}\times C_{15}$   &$u=(000000100011011100100110111001001010000000001)$,\\&&&
 $v=(110000000000010001001110100100110011100101001)$\\
 $ $&$[45,37,4]_{\rm R,C}$     &$C_{3}\times C_{15}$   &$u=(010000110101111010010111010001001110011000001)$,\\&&&
 $v=(100011011011011011100011100100011011100100011)$\\
 $ $&$[45,38,4]$     &$C_{3}\times C_{15}$   &$u=(000010010011100011011001100110010011010001110)$,\\&&&
 $v=(100001001010100111100100010111001111111010001)$\\
 $ $&$[45,39,3]$     &$C_{3}\times C_{15}$   &$u=(100011010001011010000111100001010011110011111)$,\\&&&
 $v=(101000000011110101101110110110000101110101000)$\\
\hline

 $49$&$[49,30,8]$     &$C_{7}\times C_{7}$    &$u=(011000110 1111001110000100010011011100010011001011)$,\\&&&
 $v=(0001100000000000001110011111111010011001111101100)$\\
 $ $&$[49,33,6]$     &$C_{7}\times C_{7}$    &$u=(100001101 1000111111111011010000110011000100100100)$,\\&&&
 $v=(1100000000010000000100001100010100001011011110101)$\\
 $ $&$[49,34,6]$     &$C_{7}\times C_{7}$    &$u=(000000111 1010010000010101000001010010110110100110)$,\\&&&
 $v=(0011000000000000010100000101111011110100110110011)$\\
 $ $&$[49,39,4]$     &$C_{7}\times C_{7}$   &$u=(000100110 1010000111101111010101110100010100011100)$,\\&&&
$v=(1000000000101110000001010111010010101001011010111)$\\
 $ $&$[49,40,4]$     &$C_{7}\times C_{7}$    &$u=(001100011 1001111110100100101010010100001000000101)$,\\&&&
 $v=(0110000000101011101000110110111011101011010100010)$\\
 $ $&$[49,42,4]$     &$C_{7}\times C_{7}$    &$u=(001000001 0000001000001010001001001101011101101101)$,\\&&&
 $v=(0100011000110111111110001101010001100110100011010)$\\
 $ $&$[49,43,3]$     &$C_{7}\times C_{7}$    &$u=(000010100 1000110100100000111000010001110101010011)$,\\&&&
 $v=(1000001010100000011110010010110011001101011111011)$\\
\hline

 $50$&$[50,40,4]_{\rm R}$     &$C_{5}\times C_{10}$   &$u=(000100001 00101011000110010011010111100100111111010)$,\\&&&
 $v=(10000000001000001001101100111111111011111011001001)$\\

 \hline
\end{tabular}
\end{table*}

\begin{table*}\centering
\caption{Good Checkable Codes from  $\mathbf{F}_3G$ } \label{TableF3}
\begin{tabular}[hbt]{|l|l|l|l |}
  \hline
 $ n$& Code $\mathcal{C}$      & Group $G$     &~~~~~~~~~~~~~~~~Generator Element $u$ and  Check Element $v$\\\hline\hline

%

 $20$&$[20,14,4]_{\rm R,C}$     &$C_{2}\times C_{10}$  &$u=(02101221221212221102)$,\\&&&       $v=(21010201020121202120)$\\
\hline

 $24$&$[24,18,4]_{\rm R}$     &$C_{2}\times C_{12}$   &$u=(112221001100010121120021)$,\\&&&   $v=(210202020202212102100221)$\\
 $ $&$[24,19,3]_{\rm R}$     &$C_{2}\times C_{12}$    &$u=(111120120021120102022202)$,\\&&&   $v=(100201020120222022111011)$\\
\hline

 $32$&$[32,18,8]$     &$C_{4}\times C_{8}$   &$u=(00010121101222121210121022000001)$,\\&&&   $v=(10000002000200020222002101121102)$\\
 $ $&$[32,21,6]$     &$C_{4}\times C_{8}$    &$u=(10002112121201021100202020221100)$,\\&&&    $v=(11000000000211222201002102001212)$\\
 $ $&$[32,25,4]$     &$C_{4}\times C_{8}$    &$u=(10222220211221211022021101222002)$,\\&&&    $v=(21000000021021210021212122110000)$\\
 $ $&$[32,26,4]$     &$C_{4}\times C_{8}$    &$u=(10202100101020110121210020010012)$,\\&&&    $v=(21000011221000222100220011021100)$\\
 $ $&$[32,27,3]$     &$C_{4}\times C_{8}$    &$u=(10210022020002122120010102210210)$,\\&&&   $v=(10010022011000112112002212211122)$\\
\hline

 $40$&$[40,33,4]$     &$C_{2}\times C_{20}$  &$u=(0200221122021020210111021201201122111221)$, \\&&&   $v=(1001010101011010221001011001100101222222)$\\
 $ $&$[40,34,4]$     &$C_{2}\times C_{20}$   &$u=(2200200100210120211221021102120010110101)$, \\&&&    $v=(2101020102012120102002012101210102121012)$\\
\hline

 $44$&$[44,36,4]$     &$C_{2}\times C_{22}$  &$u=(10120200202010120121001011010111022100110001)$, \\&&&        $v=(21010201020102200212212010121012101221120220)$\\
 $ $&$[44,37,4]$     &$C_{2}\times C_{22}$   &$u=(20102212121201022202000021222220222021012200)$, \\&&&       $v=(01020102010210110120102022202220221110200111)$\\
\hline

 $48$&$[48,41,4]_{\rm R}$     &$C_{4}\times C_{12}$  &{$u=(110121102110110111001110110220020112122022220001)$}, \\&&&    {$v=(210000110122221110220011120011001022112201221100)$}\\
 $ $&$[48,40,4]_{\rm R}$     &$C_{4}\times C_{12}$   &{$u=(120122000210012211010202001020120012101002100002)$}, \\&&&   {$v=(121000020001222120212221000100021210222111122221)$}~~~~~\\
 \hline
\end{tabular}
\end{table*}

\begin{table*}\centering
\caption{Good Checkable Codes from  $\mathbf{F}_4G$, where $\mathbf{F}_4=\{0,1,a,a^2=1+a\}$ } \label{TableF4}
\begin{tabular}[hbt]{ |l|l|l|l|}
  \hline
 $ n$& Code $\mathcal{C}$      & Group $G$     &~~~~~~~~~~~~~~~~~~~~~~~~~~~~~~Generator Element $u$ and Check Element $v$\\\hline\hline

 $18$&$[18,14,3]$     &$C_{3}\times C_{6}$    &$u=(a^2aa01011a^20a^2a^2aaa^2a^211)$,\\&&& $v=(a1a^2111a^2a10a^2100001a)$\\
\hline

 $25$&$[25,16,6]_{\rm R,C}$     &$C_{5}\times C_{5}$    &$u=(1a^2001a11a0a0a^2a0a^2110111a^2aa^2)$,\\&&& $v=(a1111111001a^2010a^210aa^2a100a)$\\
 $ $&$[25,19,4]_{\rm R,C}$     &$C_{5}\times C_{5}$    &$u=(01a^2001a0aaa^200a^2111a^20001a^2a0)$,\\&&& $v=(a^2111a11a^20a^20a^2a^2aa1a1a0a^2a^2000)$\\
 $ $&$[25,20,4]_{\rm R,C}$     &$C_{5}\times C_{5}$    &$u=(0a^2a^211aa^2a0aa^20a^211a0010a^2a11a^2)$,\\&&& $v=(a^211a^2a11a^2aa^2a0a^2a^201a100a^20a0a^2)$\\
 $ $&$[25,21,3]_{\rm R,C}$     &$C_{5}\times C_{5}$    &$u=(1101000aa^2aaa^2a0a^2111a^20a^20011)$,\\&&& $v=(a11a011a0aa^2a^2101a^20a^2aa1a^2a^210)$\\
\hline

 $45$&$[45,38,4]$     &$C_{3}\times C_{15}$   &$u=(0010aa^2aa^211aaa11aaa^200a01a^2010aa0100a^2a0aa00a0a00)$, \\&&   &$v=(a1111a^211a10a^2a^20a^2a^2111aa0a^20aa^2aa0aa^2a^21a00a^21a^2100a^2a^20)$\\
 $ $&$[45,39,4]$     &$C_{3}\times C_{15}$   &$u=(1a1aaaa^210a^2a^21aaaaaa^2aa^21a01a1a^21a^20a^21a0a^2a0a^211011a^20)$,      \\&&   &$v=(a^21111011a^2a00a^2a^2a11a^20aaaaa^200aa00aaa^2aa0a^211a^2a^2a00a)$\\
 $ $&$[45,34,6]$     &$C_{3}\times C_{15}$   &$u=(a^2aa0aa^2aaa^21110a01aa0111a^2aa0110a^21aa^2aa^2a^2a0a^2aa^2a1a^2a^2)$, \\&&   &$v=(1111111111100a^2a^210a011a0011aa^21a^20aa0a010a1aaa^211)$\\
 $ $&$[45,35,6]$     &$C_{3}\times C_{15}$   &$u=(1a^2111a000aa^2a^21a^2a1aaaa^21000aa^2a10aaa^21a^20a^21a^2aa1a^210a^2)$,      \\&&   &$v=(a^21111111a^211a^211aa1a000aa^2000100aa^2a1a10a^2a^20101a11)$\\
 $ $&$[45,40,3]$     &$C_{3}\times C_{15}$   &$u=(101a^211a^21aa1a1a^200a^2a^2a1aa^20a00a0a0aa^20aaa^2aaa^20a1a^21a)$,    \\&&   &{$v=(a1a^21101aa^2a^2a^20a^2a11aa^20aa0a^2a^2a^21aa^2a^200a^2a^2aa^21a1a^2a^2a1a^21a)$}\\
 $ $&$[45,41,3]$     &$C_{3}\times C_{15}$   &$u=(0a^20011a^211aa^201a1a^20a^2aa^2101a0100a^2a^2a00aa^2a0a^20aaa^21aa^2)$, \\&&   &$v=(a1a^210a^21a^2aa^20a01aa01a1a^201a0aa^20a^2110a^2a010a^21aa^201a^2a)$\\
\hline

 $49$&$[49,42,4]$     &$C_{7}\times C_{7}$    &{$u=(a^21a^2a^2011a^20111a^21a^20a^2aa^21a^2a^2a^210000
10a^2a^2aa^21a^2110aaaaa0a11a)$},\\&&   &$v=(a111a1a111a1aa1a1aa11aaaaaaa11a1aa1aa111a1a1aa111)$\\
\hline

 $50$&$[50,43,4]_{\rm R}$     &$C_{5}\times C_{10}$   &$u=(01a^2aaaaa^2a^200010a^21a^2a^20aa110a11aaa^21
a01a11a^2111a^210a1a^2110)$,       \\
&&   &$v=(a^211a^20111111a^2a^21a11111a^211a^2000000a00a1aaaaaa00a100000 )$\\
 $ $&$[50,44,4]_{\rm R}$     &$C_{5}\times C_{10}$
 &{$u=(a^2a^20a0aa1aa01aa^210a1a^21a^20a^2aa001aa^21a00a^2a^2a^210aa0001aa^2a^2a^2a)$},\\
 &&   &$v=(a1111110aa^2a1aa^201a^201aa01a^2a1aaaaaaa^2101a10a^2a0a^2a11a^2a01)$~~~~~~~~~~~~~~~~~~~~~~\,\,\\

 \hline
\end{tabular}
\end{table*}

\begin{table*}\centering
\caption{Good Checkable Codes from   $\mathbf{F}_5G$ } \label{TableF5}
\begin{tabular}[!hbt]{ |l|l|l|l|}
  \hline
 $ n$& Code $\mathcal{C}$      & Group $G$     &~~~~~~~~~~~~~~~~~~~~~~~~~~~~~~~~~Generator Element $u$ and Check Element $v$\\\hline\hline


 $18$&$[18,10,6]_{\rm R,C}$     &$C_{3}\times C_{6}$    &$u=(304442010212124112)$,\\&&& $v=(100000004013203240)$\\
 $ $&$[18,13,4]_{\rm R,C}$     &$C_{3}\times C_{6}$    &$u=(111444121401433042)$,\\&&&  $v=(100011044233322344)$\\
\hline

 $20$&$[20,15,4]_{\rm R}$     &$C_{2}\times C_{10}$   &$u=(12410122413003142121)$,\\&&& $v=(10010401103443404334)$\\
\hline

 $24$&$[24,19,4]$     &$C_{2}\times C_{12}$    &$u=(223203242014333100004101)$,\\&&& $v=(110103043433032211332104)$\\
\hline

 $32$&$[32,26,4]$     &$C_{4}\times C_{8}$    &$u=(12113331001244204302213311032203)$,\\&&& $v=(14140031004303104242220311044204)$\\
 $ $&$[32,28,3]$     &$C_{4}\times C_{8}$    &$u=(22111122403344243012322100142431)$,\\&&& $v=(41200324023142132403204113423102)$\\
\hline

 $36$&$[36,27,6]_{\rm R,C}$     &$C_{6}\times C_{6}$    &$u=(320132230330303404122130430344232343)$,      \\
 &&   &$v=(100001000434001100141404131131141404)$\\
 $ $&$\boldsymbol{[36,28,6]}^*_{\rm R,C}$        &$C_{6}\times C_{6}$    &$u=(021242402043131423014123232100132334)$,        \\
 &&   &$v=(100004000410431304002224330013242110)$\\
 $ $&$[36,30,4]_{\rm R,C}$     &$C_{6}\times C_{6}$    &$u=(430221420433120003111301342330403142)$,      \\
 &&   &$v=(100011024142020141102014233433232434)$\\
 $ $&$[36,31,4]_{\rm R,C}$     &$C_{6}\times C_{6}$    &$u=(414212431211114001024430113141242220)$,      \\
 &&   &$v=(100001244134331320112211023133431442)$\\
\hline

 $40 $&$[40,34,4]$     &$C_{2}\times C_{20}$   &$u=(1014241440444340241314221310400103102403)$,   \\
 &&   &$v=(0104010401313313423124042404242242403322)$\\
 $$&$[40,36,3]$     &$C_{2}\times C_{20}$   &$u=(3404442420430414423443124210412401010024)$,   \\
 &&   &$v=(1004042121214304324332324310211004321043)$\\
\hline

 $45$&$[45,38,4]_{\rm R}$     &$C_{3}\times C_{15}$   &$u=(422214114313301102020432222411013144100033133)$,\\
 &&   &$v=(100000011322000344433444011433222122322444122)$\\
\hline

 $48$&$[48,37,6]$     &$C_{4}\times C_{12}$   &$u=(022401214232343132104344424140031221030041132043)$,\\
 &&   &$v=(100000000003314304224422430430220301424142443412)$\\
 $ $&$[48,41,4]$     &$C_{4}\times C_{12}$ &$u=(033110404424400223213444240314124301040420320311)$,\\
 &&   &$v=(010001410400041412112313101143034044120221221020)$\\
 $ $&$[48,42,4]$     &$C_{4}\times C_{12}$   &$u=(200421304403101244441432224311111011301122004343)$,\\
 &&   &$v=(410000230302113004421310322332142021210224312422)$\\
 $ $&$[48,44,3]$     &$C_{4}\times C_{12}$   &$u=(034043422131413332341002234213011122220221033211)$,\\
 &&   &$v=(100401314123303424134222412344431004422200323034)$\\
\hline

 $72$&$\boldsymbol{[72,62,6]}^{*}$        &{$C_{6}\times C_{12}$
}&{$u=(312411232330313143111221222301122414030013401133430420133323011301020100)$}, \\
 &&   &$v=(
100000000441004102234010043124424101300211324012401114201004023203011413)$\\

 \hline
\end{tabular}
\end{table*}

 \newpage
\section{Conclusion}\label{sec:conclu}
We have studied checkable codes derived from the group ring $\mathbf{F}G$, where $\mathbf{F}$ is a finite field
and $G$ is a finite abelian group. We have introduced a notion of code-checkable group rings and determined
necessary and sufficient conditions for a group ring $\mathbf{F}G$ to be code-checkable. Based on this
characterization, we obtained two new codes which have minimum distance better than the lower bound given in
Grassl's table~\cite{Gr2010}. Various codes with minimum distance as good as the best known ones in
\cite{Gr2010} are also found. By shortening a new checkable code,  we obtain other  two  optimal  codes which
have minimum distance better than the lower bound in \cite{Gr2010}. In addition, we have proved that many
$[n,1,n]$ and $[n,n-1,2]$ MDS codes can be constructed as checkable codes. Furthermore, when $\mathbf{F}G$ is a
code-checkable  group ring, the dual of a code in $\mathbf{F}G$ may be described via a check element of the
code. This property generalizes the notions of the generator and parity-check polynomials of cyclic codes to the
multivariate case. Moreover, we have characterized the structures  of reversible  and complementary dual
checkable codes which are generalizations of reversible and complementary dual cyclic codes, respectively.

It would be interesting to study possible generalizations
of other properties of cyclic codes to this new class of codes.

\end{document}